\documentclass[twocolumn,showpacs,preprintnumbers]{revtex4}
\usepackage{amssymb}
\usepackage{amsfonts}
\usepackage{amsmath}
\usepackage{graphicx}
\usepackage{dcolumn}
\usepackage{bm}

\input{tcilatex}

\begin{document}

\title{The connection between phase synchronization in simple nonlinear
system and stationary state entanglement in its quantum counterpart}
\author{E. D. Vol}
\email{vol@ilt.kharkov.ua}
\affiliation{B. Verkin Institute for Low Temperature Physics and Engineering of the
National Academy of Sciences of Ukraine 47, Lenin Ave., Kharkov 61103,
Ukraine.}
\date{\today }

\begin{abstract}
We begin with the simple model of phase sychronization in open classical
nonlinear system which is represented in the language of angular momentum
variables. After that we propose the relevant quantum counterpart of this
system. Using the appropriate Lindblad master equation for the density
matrix of two qubit realization of such system we have revealed that
stationary state of this composite system is pure and entangled with small
dispersion of phase observable. We believe that such curious connection
between entangled stationary states of quantum composite system and phase
sychronization between its subsystems may be typical for rather wide class
of similar nonlinear open systems as well.
\end{abstract}

\maketitle

\section{Introduction}

Quantum entanglement of pure and mixed states is a notable feature of
quantum mechanics which plays a key role in all modern informational and
communicational applications of the theory, such as superdense coding,
quantum cryptography, teleportation and so on \cite{1s}. In this connection
the important question may arise: whether there is any counterpart of
entanglement for classical systems and if yes, how it can manifest itself.
This problem was studied by several authors (see for example \cite{2s-4s}),
but all of them focused attention only on closed Hamiltonian systems and
tried to find connection between the possible entanglement of ground state
of the quantum system in question and peculiarities arising in phase portret
of its classical counterpart when external parameters are changing. The main
goal of the present paper is to propose an example of a simple nonlinear
open system with phase syncronization which in addition has relevant quantum
counterpart such that stationary state of the system turns out to be pure,
entangled and reveals small dispersion of phase observable. Based on this
enlightening example one can anticipate the general attractive hypothesis
about the possible connection existing between entanglement of stationary
state in composite open quantum system and such well-known classical
phenomenon as phase synchronization of its subsystems. The present paper as
we hope may be considered as a small step on the way of confirmation of this
attractive hypothesis. So let us examine the model of phase synchronization
in nonlinear open system which can be conveniently formulated in the
language of angular momentum that interacts with its environment. The
advantage of this model is that it has the relevant quantum counterpart for
which the above mentioned connection looks especially clear. So let us
consider the angular momentum $\ \overrightarrow{L}=\left(
l_{x},l_{y},l_{z}\right) $ and assume that evolution in time of its
components is governed by the next system of equations:

\begin{eqnarray}
\frac{dl_{x}}{dt} &=&2\left( l_{y}^{2}+l_{z}^{2}\right)  \notag \\
\frac{dl_{y}}{dt} &=&-2l_{x}l_{y}  \label{1n} \\
\ \frac{dl_{z}}{dt} &=&-2l_{x}l_{z}  \notag
\end{eqnarray}%
Note some obvious features of the nonlinear model governed by the system Eq.(%
\ref{1n}). First of all it is easy to see that Eq. (\ref{1n}) implies the
relation :$\frac{dL^{2}}{dt}=0$ (where $L^{2}\equiv
l_{x}^{2}+l_{y}^{2}+l_{z}^{2}$ ) , or in other words, the total angular
moment of the system is conserved . On the other side one can assert that
independently from initial conditions $l_{x}\left( 0\right) ,l_{y}\left(
0\right) ,l_{z}\left( 0\right) $ the angular momentum under study tends to
its final stationary state with $l_{y}\left( \infty \right) =l_{z}\left(
\infty \right) =0$ and $l_{x}\left( \infty \right) =\max =\sqrt{L^{2}\left(
0\right) },$ where $L^{2}\left( 0\right) =l_{x}^{2}\left( 0\right)
+l_{y}^{2}\left( 0\right) +l_{z}^{2}\left( 0\right) .$ Let us prove this
statement. To this end in view it is convenient to write out the system Eq.(%
\ref{1n}) in so called quasithermodynamic form, namely:%
\begin{equation}
\frac{dl_{i}}{dt}=\frac{1}{2}\varepsilon _{ikl}\frac{dH}{dl_{k}}A_{l}
\label{2n}
\end{equation}%
where $\varepsilon _{ikl}$ is completely antisymmetric tensor of the third
rank, $A_{l}=\frac{1}{2}\varepsilon _{lmn}\frac{dS}{dl_{m}}\frac{dL^{2}}{%
dl_{n}}$ ; and $H\left( l_{x},l_{y},l_{z}\right) $ and $S\left(
l_{x},l_{y}l_{z}\right) $ are two given functions of the system state. In
the concrete case of the system Eq. (\ref{1n}) , $H=L^{2}$ and $S=l_{x}$.It
is easy to verify directly that Eq. (\ref{2n}) implies two general
relations, namely :1)$\frac{dH}{dt}=0,$ and 2)$\frac{dS}{dt}\geq 0.$Thus the
functions $H$ and $S$ \ governing the evolution of the system play the same
role as energy and entropy functions in standard classical thermodynamics.
As a result we arrive to the conclusion that final state of the system Eq. (%
\ref{1n}) should correspond to the state with maximal entropy function that
is with $l_{x}=\sqrt{L^{2}\left( 0\right) }$. Now we want to demonstrate
that dynamical system$\ $Eq. (\ref{1n}) may be considered as the simplest
model of phase synchronization. Remind that synchronization is a general
phenomenon of rhythm adjustment of two open nonlinear oscillating systems
that can occur at arbitrarily weak interaction between them \cite{5s}.
Further we will interested in only the case of complete phase
synchronization when the phase difference between two oscillating systems
under study tends to zero with time. Formally such situation can be
described as follows. Let us select the phases of two systems $\varphi _{1}$
and $\varphi _{2}$ and their amplitudes $r_{1}$and $r_{2}$ as dynamical
variables and write down the complete system of equations in the following
schematic form:$\frac{d\varphi _{i}}{dt}=f_{i}\left( \varphi _{1},\varphi
_{2},r_{1},r_{2}\right) $, and $\frac{dr_{i}}{dt}=g_{i}\left( \varphi
_{1},\varphi _{2},r_{1},r_{2}\right) $ where $f_{i\text{ }}$ and $g_{i}$ are
four appropriate nonlinear functions that can provide the phase
synchronization effect. One may write out this system as a system of
equations for two complex variables: $z_{1}=r_{1}e^{i\varphi _{1}}$ and $%
z_{2}=r_{2}e^{i\varphi _{2}}$. Now let us introduce three auxillary
variables $l_{x},l_{y},l_{z}$ connected with $z_{1}$ and $z_{2}$ by the next
relations:%
\begin{eqnarray}
l_{x} &=&\frac{z_{1}^{\ast }z_{2}+z_{1}z_{2}^{\ast }}{2},  \notag \\
l_{y} &=&\frac{i\left( z_{1}z_{2}^{\ast }-z_{1}^{\ast }z_{2}\right) }{2},
\label{3n} \\
l_{z} &=&\frac{\mid z_{1}\mid ^{2}-\left\vert z_{2}\right\vert ^{2}}{2}. 
\notag
\end{eqnarray}%
Using these relations one can obtain the equivalent system of evolution
equations for phase sychronization phenomenon in the language of variables $%
l_{x},l_{y},l_{z}$. Note that above mentioned transition from four variables 
$\left\{ \varphi _{i},r_{i}\right\} $ to three variables $l_{x},l_{y},l_{z}$
is the explicit analogue of the similar transformation proposed by J.
Schwinger in quantum theory. Note that in quantum mechanics such
transformation allows one to express three components of angular momentum
operator by means of four operators: $a_{1},a_{1}^{+},a_{2},a_{2}^{+}$ ,
satisfying to standard Bose commutation relations. In view of obvious
relations: $l_{x}=r_{1}r_{2}\cos \left( \varphi _{1}-\varphi _{2}\right) $
and $l_{y}=r_{1}r_{2}\sin \left( \varphi _{1}-\varphi _{2}\right) $ one can
express the condition of complete phase synchronization by two equivalent
ways :1) $l_{y}\left( \infty \right) =0$ or 2) $l_{x}\left( 0\right) =\max .$%
Thus the realization of complete phase sychronization in nonlinear model Eq.
(\ref{1n}) has been proved. Let us present now the appropriate quantum
counterpart of the proposed nonlinear classical model.The consistent way to
"quantize" classical equations of motion for open system (at least in
semiclassical approximation) was proposed by author in the paper \cite{5s},
where some instructive examples were represented as well. In the case of \
Eq. (\ref{1n}) the relevant recipe of quantization can be formulated as
follows (all necessary details the reader can find in \cite{5s}). First of
all one needs to represent the classic equations of motion for variables $%
l_{x},l_{y},l_{z}$ \ in the next specific form that allows their successive
quantization, namely:%
\begin{equation}
\frac{dl_{\alpha }}{dt}=-\left( \overrightarrow{l}\times \frac{\delta H}{%
\delta \overrightarrow{l}}\right) _{\alpha }+i\sum\limits_{k}R_{k}(%
\overrightarrow{l}\times \frac{\delta R_{k}^{\ast }}{\delta \overrightarrow{l%
}})_{\alpha }+c.c.\text{.}  \label{4n}
\end{equation}%
where $H\left( l_{i}\right) $ and $\left\{ R_{k}\left( l_{i}\right) \right\} 
$ is some given set of functions of the state, and furthermore the function $%
H\left( l_{i}\right) $ is real while $R_{k}\left( l_{i}\right) $ are complex
functions.It turns out that consistent quantum version of Eq. (\ref{4n}) can
be obtained if one by means of known classical functions $H$ and $R_{k}$
reconstructs the relevant Lindblad master equation (LME) for density matrix $%
\widehat{\rho }$ of quantum counterpart of the system of interest. Thus the
required Lindblad equation looks as :%
\begin{equation}
\frac{d\widehat{\rho }}{dt}=-i\left[ \widehat{H},\widehat{\rho }\right]
+\sum\limits_{k}\left[ \widehat{R}_{k}\widehat{\rho },\widehat{R}_{k}^{+}%
\right] +h.c.\text{.}  \label{5n}
\end{equation}%
where $\widehat{H}\left( \widehat{l_{i}}\right) ,$ $\ \widehat{R_{k}}\left( 
\widehat{l_{i}}\right) ,$ $\ \widehat{R_{k}^{+}}\left( \widehat{l_{i}}%
\right) $ \ are quantum observables that correspond to their classical
analogues in Eq.(\ref{4n}). Speaking more simply one must replace classical
variables $l_{x},l_{y,}l_{z}$ entering as arguments in functions $H,R_{k}$
by corresponding quantum operators $\widehat{l_{i}}$ that satisfy standard
commutation relations, namely: $\left[ \widehat{l_{i},}\widehat{l_{j}}\right]
=i\varepsilon _{ijk}\widehat{l_{k}}$ .\ If we return now to the concrete
model of Eq. (\ref{1n})\ \ it is easy to verify directly that the relevant
Lindblad equation for the density matrix of its quantum counterpart \ may be
represented in the next simple form:%
\begin{equation}
\frac{d\widehat{\rho }}{dt}=\left[ \widehat{R}\widehat{\rho },\widehat{R^{+}}%
\right] +\left[ \widehat{R},\widehat{\rho }\widehat{R^{+}}\right] \text{.}
\label{6n}
\end{equation}%
where the single jump operator $\widehat{R}$ has the form:$\widehat{R}=%
\widehat{l_{z}}-i\widehat{l_{y}}$ \ \ \ 

Let us now examine\ \ the LME, that is Eq. (\ref{6n}), which is
characterized by single jump operator \ \ $\widehat{R}=\widehat{l_{z}}-i%
\widehat{l_{y}}$\ \ in the simplest case of two qubit composite system for
which the density operator $\widehat{\rho }$ is given by $4\times 4$ matrix.
In standard matrix reprezentation of angular momentum operator $\widehat{L}$
for the system with $l=\frac{3}{2}$ one can write down its components as :%
\begin{eqnarray}
\widehat{l_{z}} &=&\frac{1}{2}%
\begin{pmatrix}
3 & 0 & 0 & 0 \\ 
0 & 1 & 0 & 0 \\ 
0 & 0 & -1 & 0 \\ 
0 & 0 & 0 & -3%
\end{pmatrix}%
,  \notag \\
\widehat{l_{x}} &=&\frac{1}{2}%
\begin{pmatrix}
0 & \sqrt{3} & 0 & 0 \\ 
\sqrt{3} & 0 & 2 & 0 \\ 
0 & 2 & 0 & \sqrt{3} \\ 
0 & 0 & \sqrt{3} & 0%
\end{pmatrix}%
,  \label{7n} \\
\widehat{l_{y}} &=&\frac{i}{2}%
\begin{pmatrix}
0 & -\sqrt{3} & 0 & 0 \\ 
\sqrt{3} & 0 & -2 & 0 \\ 
0 & 2 & 0 & -\sqrt{3} \\ 
0 & 0 & \sqrt{3} & 0%
\end{pmatrix}
\notag
\end{eqnarray}

Using the expressions Eq. (\ref{7n}) one can easilyobtain the required
matrix representation of jump operator $\widehat{R}$ in the next
form:\bigskip 
\begin{equation}
\widehat{R}=\widehat{l_{z}}-i\widehat{l_{y}}=\frac{1}{2}%
\begin{pmatrix}
3 & -\sqrt{3} & 0 & 0 \\ 
\sqrt{3} & 1 & -2 & 0 \\ 
0 & 2 & -1 & -\sqrt{3} \\ 
0 & 0 & \sqrt{3} & -3%
\end{pmatrix}%
\text{.}  \label{8n}
\end{equation}%
In the present paper we are interested in only the stationary solutions of
the LME,with jump operator given by Eq. (\ref{8n}). In this case it is easy
to see that all pure stationary states of the system can be found by the
solution of the equation: $\widehat{R}\left\vert \Psi \right\rangle _{st}=0$%
. Omitting the trivial algebra let us present the required result for $%
\left\vert \Psi \right\rangle _{st}$ in the form :%
\begin{equation}
\left\vert \Psi \right\rangle _{st}=\frac{1}{2\sqrt{2}}%
\begin{pmatrix}
1 \\ 
\sqrt{3} \\ 
\sqrt{3} \\ 
1%
\end{pmatrix}%
\text{.}  \label{9n}
\end{equation}%
We reveal that stationary state of our model is entangled, because it cannot
be represented in factorized form as : $\left\vert \Psi \right\rangle _{st}=%
\begin{pmatrix}
a \\ 
b%
\end{pmatrix}%
\otimes 
\begin{pmatrix}
c \\ 
d%
\end{pmatrix}%
$ where $%
\begin{pmatrix}
a \\ 
b%
\end{pmatrix}%
$ and $%
\begin{pmatrix}
c \\ 
d%
\end{pmatrix}%
$ are states of first and second qubits respectively. It is well known that
in the case of two qubit pure state \ : $\left\vert \Psi \right\rangle =$ $%
\begin{pmatrix}
\Psi _{1} \\ 
\Psi _{2} \\ 
\Psi _{3} \\ 
\Psi _{4}%
\end{pmatrix}%
$ there is the simple measure for pure state entanglement , namely, the
concurrence- $C$ . According to definition: $C=2\left\vert \Psi _{1}\Psi
_{4}-\Psi _{2}\Psi _{4}\right\vert $. It is easy to see that for the
stationary state Eq. (\ref{9n}) concurrence is equal to $\frac{1}{2}$. Now
let us clarify another intriguing point : how the stationary state of the
model Eq. (\ref{6n}) is connected with eigenstates of its quantum phase.
Here it should be noted that the concept of quantum phase operator\ in the
case of infinite dimensional Hlbert space rather complicated and is not
clear enough.This problem was discussed in quite number of papers (see the
most important contributions on this subject in the book \cite{8s}). However
for finite -dimensional quantum systems the problem may be represented much
simpler as follows: let us consider in $N$ dimensional Hilbert space $\left(
N=2l+1\text{, where }l\text{ some integer or half-integer}\right) ,$ the
raising operator $\widehat{l_{+}}$ with the next matrix elements in standard
basis $\left\vert l,m\right\rangle $ $:\left\langle l,m\right\vert \
l_{+}\left\vert l,m-1\right\rangle =\sqrt{\left( l-m+1\right) \left(
l+m\right) }.$ After that one can correctly define the polar decomposition
of operator $\widehat{l_{+}}$ in the form: $\widehat{l_{+}}=\sqrt{\widehat{%
l_{+}}\widehat{l}}e^{-i\widehat{\Phi }}.$Respectively, the lowering operator 
$\widehat{l}$ which is conjugate to operator $\widehat{l_{+}}$ \ looks as: $%
\widehat{l}=e^{i\widehat{\Phi }}\sqrt{\widehat{l_{+}}\widehat{l}\text{. }}$%
It is easy to verify directly that operators $e^{-i\widehat{\Phi }}$ and $%
e^{i\widehat{\Phi }}$ in above decompositions are unitary and all their
eigenvalues are the roots of $N$ degree from unit. For example in the case
of $N=4$, which is the main subject of our interest in present paper, above
mentioned construction may be concretized as follows. The operator $\widehat{%
l_{+}}$ has the form: $\widehat{l_{+}}=%
\begin{pmatrix}
0 & \sqrt{3} & 0 & 0 \\ 
0 & 0 & 2 & 0 \\ 
0 & 0 & 0 & \sqrt{3} \\ 
0 & 0 & 0 & 0%
\end{pmatrix}%
$ and corresponding phase operator looks as:%
\begin{equation}
e^{-i\widehat{\Phi }}=%
\begin{pmatrix}
0 & 1 & 0 & 0 \\ 
0 & 0 & 1 & 0 \\ 
0 & 0 & 0 & 1 \\ 
1 & 0 & 0 & 0%
\end{pmatrix}%
\text{.}  \label{10n}
\end{equation}%
It is easy to see by direct calculation that the above polar decompositions
are valid and furthermore $e^{i\widehat{\Phi }}.e^{-i\widehat{\Phi }}=e^{-i%
\widehat{\Phi }}.e^{i\widehat{\Phi }}=1$ that is operator $e^{-i\widehat{%
\Phi }}$ is unitary in fact.The eigenvalues $\lambda _{i}$ of operator $e^{-i%
\widehat{\Phi }\text{ }}$are equal to : $\lambda _{1}=1$, $\lambda _{2}=-1$, 
$\lambda _{3}=i$, $\lambda _{4}=-i$, and corresponding eigenvectors may be
represented as:%
\begin{eqnarray}
\left\vert 1\right\rangle &=&\frac{1}{2}%
\begin{pmatrix}
1 \\ 
1 \\ 
1 \\ 
1%
\end{pmatrix}%
;\text{ \ }\left\vert -1\right\rangle =\frac{1}{2}%
\begin{pmatrix}
1 \\ 
-1 \\ 
1 \\ 
-1%
\end{pmatrix}%
;  \label{11n} \\
\left\vert i\right\rangle &=&\frac{1}{2}%
\begin{pmatrix}
1 \\ 
i \\ 
-1 \\ 
-i%
\end{pmatrix}%
;\text{ \ }\left\vert -i\right\rangle =\frac{1}{2}%
\begin{pmatrix}
1 \\ 
-i \\ 
-1 \\ 
i%
\end{pmatrix}%
\text{.}  \notag
\end{eqnarray}%
It is easy to see that four vectors Eq. (\ref{11n}) form the complete basis
in Hilbert space of two qubit systems.Note the obvious fact that all these
states are not entangled. Now let us decompose the stationary state Eq. (\ref%
{9n})$\ $of our quantum model Eq. (\ref{6n})\ over the eigenstates of phase
operator basis Eq. (\ref{11n}).The required decomposition may be represented
as follows:%
\begin{equation}
\left\vert \Psi \right\rangle _{st}=a\left\vert 1\right\rangle +z\left\vert
i\right\rangle +z^{\ast }\left\vert -i\right\rangle \text{.}  \label{12n}
\end{equation}%
where $a$\ \ is real and $z,z^{\ast }$ \ are two conjugate complex
numbers.Comparing expressions Eq. (\ref{9n}) and Eq. (\ref{12n}) one can
find that $a=\frac{1+\sqrt{3}}{2\sqrt{2}},$ and $z=\frac{1}{4\sqrt{2}}\left(
1-\sqrt{3}\right) \left( 1+i\right) $.

Thus we arrive to the following conclusion: the entangled stationary state
of the quantum model of phase synchronization Eq. (\ref{9n}) can be
represented as superposition of three eigenstates of phase operator $e^{-i%
\widehat{\Phi }}$ with different phases. Note that the state $\left\vert
-1\right\rangle $ which corresponds to the phase $\pi $ is not included in
decomposition Eq. (\ref{12n}). It is essential that maximum contribution in
the decomposition Eq. (\ref{12n}) brings in the state with $\phi =0$. One
may easily evaluate this contribution or its relative weight $w$ as: $w=%
\frac{a^{2}}{a^{2}+2\left\vert z\right\vert ^{2}}\approx 0,9.$

Summing up the results obtained in this paper we may conclude that entangled
stationary state of the quantum counterpart of simple classical model of
phase sychronization Eq. (\ref{1n}) is very closed to the unentangled
eigenstate of phase operator with phase $\Phi =0$. Based on analysis of this
model one may anticipate the attractive hypothesis that similar relationship
may take place for a much larger class of nonlinear open systems which
demonstrate the phenomenon of phase synchronization. In this connection it
is worth noting that the possibility of similar connection on a qualitative
level have been discussed at length in known correspondence between Wolfgang
Pauli and eminent psychologist K.G. Jung \cite{8s} many years ago. The
exposition of this correspondence the reader can find in recent, rich in
content, review paper \cite{9s}. The author hopes that the simple nonlinear
model and its quantum counterpart proposed in the present paper can be
considered as the small step on the way of quantitative simulation of this
intriguing connection in more complex systems as well.

\end{document}